\documentclass[12pt,reqno]{amsart}
\allowdisplaybreaks[4]
\usepackage{graphicx} 
\usepackage{amssymb}
\usepackage{subfigure}
\usepackage{amsmath}
\usepackage{cite}
\usepackage{subfigure}
\usepackage{graphicx}
\usepackage{epstopdf}
\usepackage{lscape}
\usepackage[colorlinks,linkcolor=red,anchorcolor=blue,citecolor=blue]{hyperref}

\newtheorem{theorem}{Theorem}
\newcommand{\Cramer}[1]{\mathbf{Cramer}\left(#1\right)}
\newcommand{\diag}[1]{\mathbf{diag}\left(#1\right)}

\begin{document}
\title{The height of an  $n$th-order fundamental rogue wave for the nonlinear Schr\"odinger equation}
\author{Lihong Wang$^{1,2}$,  Chenghao Yang$^{2}$,  Ji Wang$^{1}$,  Jingsong He$^{3*}$}
\thanks{$^*$ Corresponding author: hejingsong@nbu.edu.cn, jshe@ustc.edu.cn}
\dedicatory {
$^{1}$ School of Mechanical Engineering $\&$ Mechanics, Ningbo University, Ningbo,  P.\ R.\ China \\
$^{2}$ State Key Laboratory of Satellite Ocean Environment Dynamics (Second Institute of Oceanography, SOA), P.\ R.\ China \\
$^{3}$ Department of Mathematics, Ningbo University, Ningbo, Zhejiang 315211, P.\ R.\ China}
\begin{abstract}
The height of an $n$th-order fundamental rogue wave $q_{\rm rw}^{[n]}$ for the nonlinear Schr\"odinger equation, namely $(2n+1)c$,  is proved directly by  a series of row operations on matrices appeared in the $n$-fold Darboux transformation.  Here the positive constant $c$ denotes the height of the asymptotical plane
of the rogue wave.
\end{abstract}

\maketitle \vspace{-0.9cm}

\noindent {{\bf Keywords}: Rogue wave, Nonlinear Schr\"odinger equation, Darboux transformation}

\noindent {\bf PACS} numbers: 02.30.Ik, 42.65.Tg, 42.81.Dp, 05.45.Yv  \\

\section{Introduction}

The high-intensity light is of growing importance in the  creation of  few-cycle pulses at attosecond scale \cite{brabec1,brabec2, LMreportphys}, which in general implies
the vast  boost  of the high bit-rate optical fiber communication and the birth of the  so called ``attosecond physics"\cite{krausz1}.  Parallel to above extensive researches
on this featured light from the view of the optics,  another  kind of large amplitude optical wave, namely optical rogue wave (RW) for the nonlinear Schr\"odinger(NLS) equation,
has been paid much attention since the  experimental observation at year 2011\cite{akhmediev1np}, almost 30 years later of the  discovery of the solution\cite{peregrine}.
This wave is  described vividly as  which appears from nowhere and disappears without a trace \cite{akhmediev2pla2009}.   The first-order RW, which  is also called Peregrine
soliton\cite{peregrine}, has a simple profile: two hollows allocated in the two sides of the  center peak on a non-vanishing asymptotic plane, and maximum amplitude  of the
peak is three times of the height of the plane. However, higher-order RWs have several different interesting patterns \cite{akhmediev3pla,akhmediev4pre, ling1, ling2, jkyang1,he1, matveev1non2013,matveev2thm2016,zakharovrw1}.  For example, a fundamental pattern is consisted of
a central  main peak and several gradually decreasing peaks allocated in two sides on a non-vanishing asymptotical plane.  There exists a conjecture \cite{Dubardthesis, akhmediev5pre2009,akhmediev6jpa2010}:   the height of a $n$th-order RW under the fundamental pattern is $(2n+1)$ times  of the height of the asymptotical plane, which has been confirmed for several RWs in  both theory up to  twelfth-order \cite{Gaillard1ar2016}
 and experiment results up to  fifth-order \cite{akhmediev7pre2012}. Thus, higher-order RWs have larger amplitude (larger  power), and hence can be more destructive in some disastrous events  and be more useful to generate large-intensity short optical
 pulses.  The  higher-order RWs  provide higher possibility for observation  because of their large amplitudes,  so that we can use or avoid them with more conveniences in real physical
 systems \cite{Dudley1np2014, Dudley2nc2016, Onortao1pr2013, Onortao2book2016}.  Therefore, it is physically important to pay more  attention on the height of the higher-order RWs.

  The first mathematical proof of the above expression for the maximal amplitude was given in  \cite{akhmediev6jpa2010}. Namely, the recurrence relations (7) and (8) of this work lead directly to the explicit formulae $(2n+1)$ for the expression of interest. Another work that gave the proof for the same expression for the maximal amplitude is \cite{Gaillard1ajpa2015}.
  The problem was also addressed in  \cite{chin1pla2016}.    The purpose of this paper is to provide a direct proof  of the above conjecture, which differs from above two works  given in \cite{akhmediev6jpa2010, Gaillard1ajpa2015}.  The proof of the conjecture is  a highly non-trivial work because the formula of the higher-order RW for the NLS is extremely
cumbersome such that a  fifth-order RW with eight parameters takes more than fourteen thousands pages\cite{Gaillard2ctp2015}.   The nonlinear Schr\"{o}dinger  equation\cite{chiao, zakharov1} is the form of
\begin{equation}\label{NLSeq}
iq_{t}+q_{xx}+2|q|^2q=0.
\end{equation}
Here $q=q(x,t)$ represents the envelop of electric field, $t$ is a normalized spatial variable  and $x$ is a normalized  time variable.  In optics, the squared modulus $|q|^2$ usually denotes a measurable quantity optical power (or intensity).  The NLS equation is a widely applicable integrable system \cite{ablowitzbook} in physics, which is solved by several methods such as
the inverse scattering method \cite{ablowitzbook}, the Hirota method \cite{hirotabook} and the Darboux transformation(DT) \cite{matveev1book1991}.  Recently, the height of multi-breather of the NLS has been given in references \cite{chin1pla2016, chin2arxiv2016},  but which can not imply the height of the RWs because of the appearing of an indeterminate form $\frac{0}{0}$ under the
degeneration of eigenvalues, namely $a_i\rightarrow \frac{1}{2}$ in reference \cite{chin1pla2016} or  $\nu_i\rightarrow 1$ in reference \cite{chin2arxiv2016}.
In general,  this  indeterminate form is unavoidable to construct RWs for many  equations\cite{akhmediev8pre2012pre2ndabs, akhmedie9vpra2013preclassifying,he2,he3,he4,he5,he6,he7,he8,he9,he10,he11,he12,he13,cheny, degasperis1, heguoprsa, chensh, qinzy, zhao1,zhao2} by the DT.  Thus it is worthwhile to provide a direct proof of the above conjecture based on the determinant representation of the  DT for the NLS\cite{matveev1book1991,he1,hedarbroux}.

The rest of the paper is organized as follows. After  a brief summary of the $n$th-order breather, a new formula of the $n$th-order RW $q^{[n]}_{\rm rw}(x,t)$ is given  by using
a newly introduced function {\bf Cramer}  in section II. In section III, we provide a direct proof of the conjecture on the height of  the  $q^{[n]}_{\rm rw}(x,t)$  by a series of row operations on the corresponding matrix appeared in the $n$-fold DT .

\section{The $n$th-order Rogue Waves of the  NLS generated by the $n$-fold DT}
We recall briefly to construct  higher-order RWs  from  higher-order breathers of the NLS by the DT\cite{he1}. In order to construct the DT, it is necessary to introduce a proper ``seed" as follows:
\begin{equation}\label{q0}
q^{[0]} =c e^{i \rho},
\end{equation}
with $\rho= a x + \left( 2c^2-a^2 \right) t,  a \in \mathbb{R}, c> 0$, $i^2=-1$.   Corresponding to this ``seed" solution,  the eigenfunction of the Lax equation associated with $\lambda$
is expressed by
\begin{equation}\label{phi}
\phi(\lambda)=
\left(
\begin{array}{ll}
 e^{ i  \frac{\rho}{2}  }  & 0 \\
0 & e^{-i \frac{\rho}{2}   }
\end{array}
\right)
\varphi(\lambda),
\end{equation}
\begin{equation}\label{varphi}
\varphi(\lambda)=
\left(
\begin{array}{ll}
\varphi_1(\lambda)   \\
\varphi_2(\lambda)
\end{array}
\right)=
\left(
\begin{array}{ll}
c e^{ i  d(\lambda)}   +i \left( \lambda +\frac{a}{2}+ h(\lambda) \right) e^{ -i d(\lambda) }   \\
c e^{-i     d(\lambda)}  +i \left( \lambda +\frac{a}{2} + h(\lambda) \right) e^{i d(\lambda) }
\end{array}
\right),
\end{equation}
in which  $ h(\lambda)=\sqrt{\left( \lambda+\frac{a}{2} \right)^2 + c^2}$, $d(\lambda) =(x+(2\lambda-a)t)   h(\lambda)$.  In order to construct $n$th-fold DT of the NLS,
  introduce following  $2n$ eigenfunctions
\begin{equation}\label{sec1}
f_{2k-1}\triangleq \phi(\lambda_{2k-1}) =
\left(
\begin{array}{ll}
f_{2k-1,1}\\
f_{2k-1,2}
\end{array}
\right)\;  \text{for }  \lambda_{2k-1},
\end{equation}
and
\begin{equation}\label{sec2}
 f_{2k}\triangleq
\left(
\begin{array}{rr}
-f_{2k-1,2}^*\\
f_{2k-1,1}^*
\end{array}
\right)=
\left(
\begin{array}{ll}
f_{2k,1}\\
f_{2k,2}
\end{array}
\right)\;  \text{for  }  \lambda_{2k}= \lambda_{2k-1}^*,
\end{equation}
and $k=1,2,3,\cdots, n$.  Here the asterisk denotes the complex conjugation. Using above ``seed" solution and the eigenfunctions, the $n$-fold DT generates an $n$th-order  breather of the NLS \cite{he1}
\begin{equation}\label{qn}
q^{[n]}=q^{[0]}-2i\frac{|\Delta_1^{[n]}|}{|\Delta_2^{[n]}|}.
\end{equation}
Here, $|\cdot|$ denotes the determinant of a matrix, and two matrices in $q^{[n]}$ are

\begin{equation}
\begin{array}{l}
\Delta_1^{[n]}=(\alpha_1,\alpha_2,\cdots,\alpha_{2n-1},\alpha_{2n+1}), \\
\Delta_2^{[n]}=(\alpha_1,\alpha_2,\cdots,\alpha_{2n-1},\alpha_{2n}),
\end{array}
\end{equation}
with
\begin{equation}\label{alpha2k-1}
\alpha_{2k-1}=
\left(
\begin{array}{c}
\lambda_1^{k-1} f_{1,1} \\
\lambda_2^{k-1} f_{2,1}  \\
\vdots \\
\lambda_{2n}^{k-1} f_{2n,1}
\end{array}
\right)=
\left(
\begin{array}{cc}
\lambda_1^{k-1} \varphi_{1,1} \\
\lambda_2^{k-1} \varphi_{2,1}  \\
\vdots \\
\lambda_{2n}^{k-1} \varphi_{2n,1}
\end{array}
\right)
e^{ i  \frac{\rho}{2}  }, \,
(k=1,2,\cdots,n+1),
\end{equation}

and

\begin{equation}\label{alpha2k}
\alpha_{2k}=
\left(
\begin{array}{c}
\lambda_1^{k-1} f_{1,2} \\
\lambda_2^{k-1} f_{2,2}  \\
\vdots \\
\lambda_{2n}^{k-1} f_{2n,2}
\end{array}
\right)=
\left(
\begin{array}{c}
\lambda_1^{k-1} \varphi_{1,2} \\
\lambda_2^{k-1} \varphi_{2,2}  \\
\vdots \\
\lambda_{2n}^{k-1} \varphi_{2n,2}
\end{array}
\right)
e^{ -i  \frac{\rho}{2}  }, \,
(k=1,2,\cdots,n).
\end{equation}
It can be seen that  the $n$th-order breather $q^{[n]}$ has two variables $x$ and $t$,  two real parameters $a$ and $c$,  and $n$ unique   complex spectrum
parameters $\lambda_k$  corresponding  eigenfunctions $f_{k}(k=1,3, 5, \cdots,2n-1)$.

To be more convenient to  formulate the $n$th-order breather,  introduce $\mathbf{Cramer(\cdot)}$ function,
\begin{equation}\label{cramer}
\Cramer{\Delta}=\frac{|\Delta_1|}{|\Delta_2|}.
\end{equation}
Here,  $\Delta=(\vec{a}_1,\vec{a}_2,\cdots,\vec{a}_{2n}|\vec{a}_{2n+1})$ is an augmented matrix  of $2n$-dimensional column vectors $\vec{a}_j(j=1,2,3,\cdots, 2n+1)$, $\Delta_1=(\vec{a}_1,\vec{a}_2,\cdots,\vec{a}_{2n-1},\vec{a}_{2n+1})$ and $\Delta_2=(\vec{a}_1,\vec{a}_2,\cdots,\vec{a}_{2n-1},\vec{a}_{2n})$ are two sub-matrices
 of  $\Delta$.
Setting
\begin{equation}\label{delta3}
\Delta^{[n]}=(\alpha_1,\alpha_2,\cdots,\alpha_{2n-1},\alpha_{2n}|\alpha_{2n+1})=\Delta^{[n]}(f_{j1}, f_{j2};\lambda_j),
\end{equation}
in which the second equality just means that $\Delta^{[n]}$ is generated by functions $f_j(j=1,3,5,\cdots,$ \\ $2n-1)$.  Note that $f_{2k}$ and  $\lambda_{2k}(k=1,2,3,\cdots,n)$ are given by  (\ref{sec2}). Using newly introduced {\bf Cramer} function, then $n$th-order breather (\ref{qn}) can  be re-written as
\begin{equation}\label{qncr}
 q^{[n]}=q^{[0]}-2i  \Cramer{\Delta^{[n]} },
 \end{equation}
which will be used to study the height of  the $n$th-order RWs.

Simplify determinants in numerator and denominator of  $\Cramer{\Delta^{[n]}}$ simultaneously by using (\ref{alpha2k-1}) and (\ref{alpha2k}), then
 \begin{equation}\label{cramerRowP}
\Cramer{\Delta^{[n]}(f_{j1}, f_{j2};\lambda_j)}=\Cramer{\Delta^{[n]}(\varphi_{j1}, \varphi_{j2};\lambda_j)} \cdot e^{i\rho}.
\end{equation}
 Equation (\ref{cramerRowP}) shows that   $\Cramer{\Delta^{[n]}}$ is generated equivalently
by functions  $\varphi_{j}(j=1,3,5,$ \\ $\dots,2n-1)$ except of a phase $\rho$.  Moreover, the  {\bf Cramer} function  has  an important property. \\
{\bf Lemma 1} Set  $P$ is an invertible $2n\times 2n$ matrix, then
\begin{equation}\label{cramerP}
\Cramer{P \cdot \Delta}=\Cramer{\Delta}.
\end{equation}
\noindent {\bf Proof:}
\begin{equation*}
\Cramer{P \cdot \Delta}=\frac{| P  \Delta_{1} |  }{ | P \Delta_{2}  |  }=\frac{|P| |  \Delta_{1}  |  }{ | P| | \Delta_{2}  |  }=\frac{ | \Delta_{1}  |  }{  | \Delta_{2}  |  }=\Cramer{\Delta}.
\qed
\end{equation*}

\noindent
This lemma shows that the ratio of  the two determinants in $\Cramer{P \cdot \Delta}$ is invariant under  elementary row operations, and will be used repeatedly to do row operations on the matrices appeared in the $n$th-order RW.

It is known that  an $n$th-order RW $q_{\rm rw}^{[n]}$ of the NLS  is obtained from an  $n$th-order breather $q^{[n]}$ (\ref{qncr}) by higher-order Taylor expansion of an indeterminate form $\frac{0}{0}$ which is appeared from  the  double degeneration $\lambda_j\rightarrow\lambda_1 \rightarrow \lambda_0=-\frac{a}{2}+ i c(j=1,3,5, \cdots, 2n-1)$ \cite{he1}. According to (\ref{qncr}) and (\ref{cramerRowP}), the $n$th-order RW of the NLS becomes
\begin{equation}\label{rw2}
q^{[n]}_{\rm rw}=q^{[n]}_{\rm rw}(x,t)=q^{[0]}-2i  \Cramer{\Delta^{[n]}_{\rm rw}  } \cdot e^{i\rho}.
\end{equation}
Here, matrix $\Delta^{[n]}_{\rm rw}$ is  defined by following elements
\begin{equation}\label{delta_rw_jk}
\left(\Delta^{[n]}_{\rm rw} \right)_{j,k}=
\frac{1}{[\frac{j+1}{2}]!} \frac{ \partial^{[\frac{j+1}{2}]} } {\partial \sqrt{\epsilon} }  (  \Delta^{[n]}(\varphi_{11},\varphi_{12};\lambda_0+\epsilon )   )_{j,k} |_{\epsilon=0 },
\end{equation}
because of
\begin{equation}\label{delta_rw_11}
\frac{1}{[\frac{j+1}{2}]!} \frac{ \partial^{[\frac{j+1}{2}]} } {\partial \sqrt{\epsilon} }  (  \Delta^{[n]}(\varphi_{j1},\varphi_{j2};\lambda_0+\epsilon )   )_{j,k} |_{\epsilon=0 }=
\frac{1}{[\frac{j+1}{2}]!} \frac{ \partial^{[\frac{j+1}{2}]} } {\partial \sqrt{\epsilon} }  (  \Delta^{[n]}(\varphi_{11},\varphi_{12};\lambda_0+\epsilon )   )_{j,k} |_{\epsilon=0 },
\end{equation}
under the  double degeneration of eigenvalues $\lambda_j\rightarrow\lambda_1=\lambda_0+\epsilon$.
It can be seen clearly  that    $\Delta^{[n]}_{\rm rw}$ is just generated by one eigenfunction $f_1$, or equivalently by two components $\varphi_{11}$ and $\varphi_{12}$ of $\varphi_1$.
By the simplification of  the formula of $\Delta^{[n]}_{\rm rw}$,  then
\begin{equation}\label{delta_rw_12}
\Delta^{[n]}_{\rm rw}=\Delta^{[n]}_{\rm rw}(\varphi_{11},\varphi_{12}),
\end{equation}
\begin{equation}\label{phi1112ij}
\left(\Delta^{[n]}_{\rm rw}(\varphi_{11},\varphi_{12})  \right)_{j,k}=
\left\{
\begin{array}{ll}
\frac{1}{(\frac{j+1}{2})!} \frac{\partial^{\frac{j+1}{2}}}{\partial \sqrt{\epsilon}}   (  (\lambda_0+\epsilon)^{\frac{k-1}{2}} \varphi_{11}  )|_{\epsilon=0}, & j\in {\rm Odd}, k\in {\rm Odd},  \\
\frac{1}{(\frac{j+1}{2})!} \frac{\partial^{\frac{j+1}{2}}}{\partial \sqrt{\epsilon}}   (  (\lambda_0+\epsilon)^{\frac{k-2}{2}} \varphi_{12}  )|_{\epsilon=0}, & j\in {\rm Odd}, k\in {\rm Even}, \\
\frac{1}{(\frac{j}{2})!} \frac{\partial^{\frac{j}{2}}}{\partial \sqrt{\epsilon}}   ( - (\lambda_0+\epsilon)^{\frac{k-1}{2}} \varphi_{12}  )^*|_{\epsilon=0},  & j\in {\rm Even}, k\in {\rm Odd}, \\
\frac{1}{(\frac{j}{2})!} \frac{\partial^{\frac{j}{2}}}{\partial \sqrt{\epsilon}}    (  (\lambda_0+\epsilon)^{\frac{k-2}{2}} \varphi_{11}  )^*|_{\epsilon=0},  & j\in {\rm Even}, k\in {\rm Even}.
\end{array}
\right.
\end{equation}
It it clear that the coefficient of $\epsilon^{l}$ in above formula has zero contribution  when $l$ is an integer.
Formula (\ref{rw2}) of the  $n$th-order RW $q^{[n]}_{\rm rw}(x,t)$   is crucial to study the height in this paper, because it is convenient to introduce row operations in order to simplify determinants in numerator and denominator.

\section{ The Height of an $n$th-order fundamental RW}
The $q^{[n]}_{\rm rw}(x, t)$ in (\ref{rw2}) is an  $n$th-order fundamental rogue wave of the NLS\cite{he1}, and the height of its asymptotical plane is $c$.   Because one can always move the central peak to origin of the coordinate on the $(x,t)$-plane, we shall set $x=0$ and $t=0$ in $q^{[n]}_{\rm rw}$ in following theorem to study the height  without the loss of the generality.
\begin{theorem}
The height of an $n$th-order fundamental RW is $\left| q^{[n]}_{\rm rw} \right|_{\rm height} = (2n+1) c $. Here $n$ is a positive integer and $c$ is the height of the asymptotical plane.
\end{theorem}
\noindent {\bf Proof:} We shall prove the theorem by four steps from  $q^{[n]}_{\rm rw}(x,t)$ in  (\ref{rw2}).  The main idea of the calculation of
$\Cramer{\Delta^{[n]}_{\rm rw}}$ is to utilize row operations according to {\bf Lemma 1},  such that the matrix $\Delta^{[n]}_{\rm rw}$ becomes a strict upper triangular matrix.

\subsection*{Step 1: Simplify the formula of $q^{[n]}_{\rm rw}(0,0)$}
It is known by double  degeneration of eigenvalues that there is only one eigenfunction  $\varphi_1$ in  $q^{[n]}_{\rm rw}(x,t)$ associated with eigenvalue $\lambda_1=\lambda_0+\epsilon$,
then
\begin{equation}\label{hformula}
h=h(\lambda_1)=\sqrt{(ic+\epsilon)^2+c^2}=\sqrt{2ic\epsilon+\epsilon^2}=h_0  \sqrt{\epsilon} \sqrt{2ic},
\end{equation}
 here,
\begin{equation}\label{h0exp}
h_0=\sqrt{1+\frac{\epsilon}{2ic}}= \sum_{k=0}^{\infty} c_k \epsilon^k=c_0+c_1 \epsilon + c_2 \epsilon^2 + \cdots + c_k \epsilon^k + \cdots,
\end{equation}
with Taylor expansion coefficients
\begin{equation}\label{ck}
c_0=1, c_1= \frac{ 1  }{ 2  \cdot(2ic)  }, c_k= \frac{ (-1)^{k-1} \cdot 1\cdot 3 \cdot 5 \cdots (2k-3)   }{2^k \cdot (2ic)^k \cdot k!}, (k=2,3,\cdots).
\end{equation}
Setting  $x=0$ and $t=0$ in $q^{[n]}_{\rm rw}$ (\ref{rw2}), yields
\begin{equation}\label{f11epsilon2}
\left\{
\begin{array}{ll}
\varphi_{11} &= i (\epsilon + h),  \\
\varphi_{12} &= i (\epsilon + h),
\end{array}
\right.
\end{equation}
and then
\begin{equation}\label{rw3}
q_{\rm rw}^{[n]}(0,0)=c-2i\Cramer{\Delta_{\rm rw}^{[n]}(i (\epsilon +   h),i (\epsilon + h))  }.
\end{equation}
Because  coefficient of $\epsilon^{l}$ in the calculation of $\Delta^{[n]}_{\rm rw}$ through (\ref{phi1112ij})  has zero contribution when $l$ is an integer,
the formula of the $n$th-order RW is further simplified as
\begin{equation}\label{rw4}
q_{\rm rw}^{[n]}(0,0)=c-2i\Cramer{\Delta_{\rm rw}^{[n]}(ih, ih))  }.
\end{equation}

\subsection*{Step 2: Remove the common factors in each row of $\Delta_{\rm rw}^{[n]}$}

According to (\ref{cramerP}) in {\bf Lemma 1},  the elementary row operations $P_0$ for $\Delta_{\rm rw}^{[n]}(ih, ih))$
 remove a  nonzero  common factor $i  \sqrt{2ic} $ in odd rows while  $-i \sqrt{-2ic}$  in  even rows, and  the
transformed matrix $\Delta_{\rm rw}^{[n]}(h_0\sqrt{\epsilon} ,h_0\sqrt{\epsilon} )$ is given in (\ref{delta3simp2}). This implies

\begin{equation}\label{f11simp2}
\Cramer{\Delta_{\rm rw}^{[n]}(ih,ih)} =\Cramer{ P_0   \Delta_{\rm rw}^{[n]}(ih,ih)}  = \Cramer{\Delta_{\rm rw}^{[n]}(h_0\sqrt{\epsilon} ,h_0\sqrt{\epsilon} )},
\end{equation}
and the second equality can be seen from (\ref{hformula}).   Here
\begin{equation*}\label{P0}
P_0=\diag{r_1,r_2,r_1,r_2  ,\cdots, r_1,r_2 },
\end{equation*}
$r_1=(i\sqrt{2ic})^{-1}$ and $\quad r_2=(-i\sqrt{-2ic})^{-1}$.  Therefore  the $n$th-order RW becomes
\begin{equation}\label{rw5}
q_{\rm rw}^{[n]}(0,0)=c-2i\Cramer{\Delta_{\rm rw}^{[n]}(h_0\sqrt{\epsilon}, h_0\sqrt{\epsilon} )}.
\end{equation}
\subsection*{Step 3: Transform $\Delta_{\rm rw}^{[n]}$ to be a block upper triangular matrix }
We further construct a series of  row operation matrices $P_1,P_2, \cdots, P_{n-1}$(see appendix) acting
on $\Delta_{\rm rw}^{[n]}(h_0\sqrt{\epsilon} ,h_0\sqrt{\epsilon} )$  such that the transformed matrix

$$
\Delta_{\rm rw}^{[n]}( \sqrt{\epsilon} , \sqrt{\epsilon} )=P_{n-1} \cdots P_2 P_1\Delta_{\rm rw}^{[n]}(h_0\sqrt{\epsilon} ,h_0\sqrt{\epsilon} )
$$
 is   given by (\ref{delta3simp3}). According to {\bf Lemma 1},
 $$
\Cramer{\Delta_{\rm rw}^{[n]}(h_0\sqrt{\epsilon} ,h_0\sqrt{\epsilon} )}=
\Cramer{  \Delta_{\rm rw}^{[n]}( \sqrt{\epsilon} , \sqrt{\epsilon} )},
$$
which leads to  a new formula of the $n$th-order RW
\begin{equation}\label{rw6}
q_{\rm rw}^{[n]}(0,0)=c-2i\Cramer{\Delta_{\rm rw}^{[n]}(\sqrt{\epsilon}, \sqrt{\epsilon} )}.
\end{equation}

\subsection*{Step 4: Transform $\Delta_{\rm rw}^{[n]}$ to be a strict upper triangular matrix }
We are now in a position to do final row operations on the matrix in {\bf Cramer}, i.e. $P_n \Delta_{\rm rw}^{[n]}(\sqrt{\epsilon}, \sqrt{\epsilon} )$,
such that matrix given by (\ref{delta3simp3})   becomes a strict  upper triangular  matrix, and thus the {\bf Cramer} of the transformed matrix is $nic$. Here the $2n\times2n$ block diagonal matrix $P_n$ is
given by
\begin{equation}\label{Pn}
P_{n}=\diag{J,J,\cdots,J},
\end{equation}
and
\begin{equation}\label{J}
J=\left(
\begin{array}{ll}
1 & 0   \\
1 & 1
\end{array}
\right).
\end{equation}
According to {\bf Lemma 1}, $\Cramer{\Delta_{\rm rw}^{[n]}(\sqrt{\epsilon}, \sqrt{\epsilon} )}=nic$. Substitute it back into
formula (\ref{rw6}), then
\begin{equation} \label{rw7}
q^{[n]}_{\rm rw}(0,0)=c-2i(nic)=(2n+1) c.
\end{equation}
Therefore, the  height of an $n$th-order fundamental RW of the NLS is
\begin{equation}\label{maxheightnrw}
\left| q^{[n]}_{\rm rw} \right|_{\rm height}= \left| q^{[n]}_{\rm rw}(0,0) \right| = (2n+1) c.
\end{equation}
This is the end of the proof.  $\qed$

{\bf Acknowledgments}  {\noindent \small
\textbf{Acknowledgments.}
This work is supported by the NSF of China under Grant No.11671219,  and the K.C. Wong Magna Fund in Ningbo University.
This study is also supported by the open Fund of the State Key Laboratory of Satellite Ocean Environment Dynamics, Second Institute of Oceanography(No. SOED1708).


\clearpage

\section*{Appendix}
A series of row operation matrices in {\bf Step 3:}

\begin{equation}\label{P1}
P_1=\left(
\begin{array}{cclllllll}
1    & 0      & 0 & 0 & 0 & 0& \cdots &0& 0  \\
0    & 1      & 0 & 0 & 0 & 0& \cdots &0& 0  \\
-c_1 & 0      & 1 & 0 & 0 & 0& \cdots &0& 0  \\
0    & -c_1^* & 0 & 1 & 0 & 0& \cdots &0& 0  \\
-c_3 & 0      & 0 & 0 & 1 & 0 & \cdots &0& 0  \\
0    & -c_3^* & 0 & 0 & 0 & 1 & \cdots &0& 0  \\
\vdots & \vdots & & & & & \ddots \\
-c_{m-1} & 0      & 0 & 0 & 0 & 0 & \cdots &1& 0  \\
0    & -c_{m-1}^* & 0 & 0 & 0 & 0 & \cdots &0& 1
\end{array}
\right)
\end{equation}

\[\]

\begin{equation}\label{P2}
P_2=\left(
\begin{array}{llcclllll}
1 & 0 & 0   & 0 & 0 & 0& \cdots &0& 0  \\
0 & 1 & 0   & 0 & 0 & 0& \cdots &0& 0  \\
0 & 0 & 1   & 0      & 0 & 0& \cdots &0& 0  \\
0 & 0 & 0   & 1      & 0 & 0& \cdots &0& 0  \\
0 & 0 &-c_1 & 0      & 1 & 0& \cdots &0& 0  \\
0 & 0 & 0   & -c_1^* & 0 & 1 & \cdots &0& 0  \\
0 & 0 &-c_3 & 0      & 0 & 0 & \cdots &0& 0  \\
0 & 0 & 0   & -c_3^* & 0 & 0 & \cdots &0& 0  \\
  &   & \vdots & \vdots & & & \ddots \\
0 & 0 &-c_{m-2} & 0      & 0 & 0 & \cdots &1& 0  \\
0 & 0 & 0   & -c_{m-2}^* & 0 & 0 & \cdots &0& 1
\end{array}
\right)
\end{equation}

$$\cdots$$

\begin{equation}\label{Pn_1}
P_{n-1}=\left(
\begin{array}{lllllccll}
1 & 0 & \cdots & 0 & 0 & 0 & 0&0& 0  \\
0 & 1 & \cdots & 0 & 0 & 0 & 0&0& 0  \\
  &   &  \ddots & & & \vdots & \vdots \\
0 & 0 & \cdots & 1 & 0 & 0 & 0&0& 0  \\
0 & 0 & \cdots & 0 & 1 & 0 & 0&0& 0  \\
0 & 0 & \cdots & 0 & 0 & 1 & 0 &0& 0  \\
0 & 0 & \cdots & 0 & 0 & 0 & 1 &0& 0  \\
0 & 0 & \cdots & 0 & 0 & -c_1 & 0      &1& 0  \\
0 & 0 & \cdots & 0 & 0 & 0    & -c_1^* &0& 1
\end{array}
\right)
\end{equation}

\begin{landscape}
The transformed matrix after {\bf Step 2}:

\[\]
{
\footnotesize
\begin{equation}\label{delta3simp2}
\left(
\begin{array}{cc cc c cc|c}
1  & 1 &   \lambda_0   & \lambda_0     &   \cdots &    \lambda_0^{n-1}         & \lambda_0^{n-1}       &    \lambda_0^{n}  \\
-1 & 1 &  -\lambda_0^* & \lambda_0^*   &   \cdots &     -{\lambda_0^*}^{n-1}   & {\lambda_0^*}^{n-1}   &    -{\lambda_0^*}^{n}   \\
c_1    & c_1   &   \lambda_0 c_1 +  1    &  \lambda_0 c_1 + 1       &  \cdots &    \lambda_0^{n-1} c_{1} + \binom{n-1}{1}  \lambda_0^{n-2} & \lambda_0^{n-1} c_{1} + \binom{n-1}{1}  \lambda_0^{n-2}  &  \lambda_0^{n} c_{1} + \binom{n}{1}  \lambda_0^{n-1}    \\
-c_1^* & c_1^* & - \lambda_0^* c_1^* - 1 &  \lambda_0^*  c_1^* + 1  &  \cdots &   -{\lambda_0^*}^{n-1} c_{1}^* - \binom{n-1}{1} {\lambda_0^*}^{n-2}  &  {\lambda_0^*}^{n-1} c_{1}^* + \binom{n-1}{1} {\lambda_0^*}^{n-2} &    -{\lambda_0^*}^{n} c_{1}^* - \binom{n}{1}  {\lambda_0^*}^{n-1} \\
\vdots  &  \vdots & \vdots & \vdots &  \ddots &   \vdots & \vdots &   \vdots \\
c_{n-1}     & c_{n-1}   &    \lambda_0 c_{n-1} + c_{n-2} &  \lambda_0 c_{n-1} + c_{n-2}    & \cdots &  \sum_{j=0}^{n-1} \binom{n-1}{j} \lambda_0^{n-j-1} c_{n-j-1} & \sum_{j=0}^{n-1} \binom{n-1}{j} \lambda_0^{n-j-1} c_{n-j-1}  & \sum_{j=0}^{n-1} \binom{n}{j} \lambda_0^{n-j} c_{n-j-1} \\
-c_{n-1}^*  & c_{n-1}^* &   -\lambda_0^* c_{n-1}^* - c_{n-2}^* &  \lambda_0^* c_{n-1}^* + c_{n-2}^*   &\cdots &  - \sum_{j=0}^{n-1} \binom{n-1}{j} {\lambda_0^*}^{n-j-1} c_{n-j-1}^*  &    \sum_{j=0}^{n-1} \binom{n-1}{j} {\lambda_0^*}^{n-j-1} c_{n-j-1}^*   & - \sum_{j=0}^{n-1} \binom{n}{j} {\lambda_0^*}^{n-j} c_{n-j-1}^*
\end{array}
\right)
\end{equation}
}

\[\]

The transformed matrix after {\bf Step 3}:

\begin{equation}\label{delta3simp3}
\left(
\begin{array}{cc cc cc c cc|c}
1 & 1 & \lambda_0   & \lambda_0   &  \lambda_0^2   & \lambda_0^2  & \cdots &  \lambda_0^{n-1}  & \lambda_0^{n-1}  &   \lambda_0^{n}  \\
-1 & 1  &  -\lambda_0^* & \lambda_0^*    &   -{\lambda_0^*}^2   & {\lambda_0^*}^2   & \cdots &    -{\lambda_0^*}^{n-1}   & {\lambda_0^*}^{n-1}   &   -{\lambda_0^*}^{n}   \\
0 & 0  & 1 & 1 &  2\lambda_0  & 2\lambda_0  &\cdots &   \binom{n-1}{n-2}  \lambda_0^{n-2} & \binom{n-1}{n-2}  \lambda_0^{n-2}  &  \binom{n}{n-1}    \lambda_0^{n-1}  \\
0 & 0  &  - 1 &  1  &   -2\lambda_0^*   & 2\lambda_0^*    &\cdots &  -\binom{n-1}{n-2}  {\lambda_0^*}^{n-2}  &  \binom{n-1}{n-2}  {\lambda_0^*}^{n-2} &  - \binom{n}{n-1} {\lambda_0^*}^{n-1}  \\
0 & 0  &   0 &  0  & 1  & 1   &\cdots &    \binom{n-1}{n-3} {\lambda_0}^{n-3} & \binom{n-1}{n-3} {\lambda_0}^{n-3}   &  \binom{n}{n-2} {\lambda_0}^{n-2} \\
0 & 0  &  0 &  0   &   -1   & 1   &\cdots &    - \binom{n-1}{n-3}  {\lambda_0^*}^{n-3}  & \binom{n-1}{n-3} {\lambda_0^*}^{n-3}  &   -\binom{n}{n-2} {\lambda_0^*}^{n-2}  \\
\vdots  &  \vdots  & \vdots &  \vdots &  \vdots &  \vdots & \ddots &  \vdots & \vdots &  \vdots \\
0 & 0 &  0 &  0  & 0 &  0 & \cdots & 1 &1&    \binom{n}{1} \lambda_0 \\
0 & 0 &  0 &  0  & 0 &  0 & \cdots &-1 &1&  - \binom{n}{1}  \lambda_0^*
\end{array}
\right)
\end{equation}

\end{landscape}

\end{document}